\documentclass[aps,twocolumn,superscriptaddress,pre]{revtex4-2} \usepackage{amssymb,amsmath,amsthm,amsfonts,amsbsy,mathrsfs}
\usepackage{xcolor,hyperref}
\hypersetup{
   colorlinks,
   linkcolor={blue!50!black},
   citecolor={blue!50!black},
   urlcolor={blue!80!black}
}
\usepackage{graphicx}
\usepackage{color}
\usepackage{bm}% bold math
\usepackage[english]{babel}

\newcommand{\obp}{\omega_{\rm bp}}

\usepackage{physics}
\begin{document}

\title{Geometrical and Vibrational Properties of the Defects Driving the Boson Peak}

\author{Shivam Mahajan}
\affiliation{Division of Physics and Applied Physics, School of Physical and
Mathematical Sciences, Nanyang Technological University, Singapore}
\author{Darryl Seow Yang Han}
\affiliation{Division of Physics and Applied Physics, School of Physical and
Mathematical Sciences, Nanyang Technological University, Singapore}
\author{Cunyuan Jiang}
\affiliation{Wilczek Quantum Center, Shanghai Jiao Tong University, Shanghai 200240, China.}
\affiliation{Shanghai Research Center for Quantum Sciences, Shanghai 201315, China.}
\affiliation{Department of Fundamental Engineering, Institute of Industrial Science, The University of Tokyo, 4-6-1 Komaba, Meguro-ku, Tokyo 153-8505, Japan.}
\author{Matteo Baggioli}
\email{b.matteo@sjtu.edu.cn}
\affiliation{Wilczek Quantum Center, Shanghai Jiao Tong University, Shanghai 200240, China.}
\affiliation{Shanghai Research Center for Quantum Sciences, Shanghai 201315, China.}
\author{Massimo Pica Ciamarra}
\email{massimo@ntu.edu.sg}
\affiliation{Division of Physics and Applied Physics, School of Physical and
Mathematical Sciences, Nanyang Technological University, Singapore}
\affiliation{Consiglio Nazionale delle Ricerche, CNR-SPIN, Napoli, I-80126, Italy}

\date{\today}% It is always \today, today,

\begin{abstract}
In amorphous solids, the vibrational density of states shows an excess of modes over the Debye model, known as the boson peak, whose origin remains unclear. Studies suggest a link to quasi-localized nonphononic vibrations or `defects,' but identifying them is challenging due to hybridization with phonons that renders methods based on localization properties, such as the participation ratio, unreliable. We introduce a practical method to separate hybridized phonons from localized vibrations and find that boson peak phonons hybridize with compact, two-dimensional defects exhibiting oscillatory pure shear deformations. 
These two-dimensional defects are also exposed by the procedure recently employed to identify stringlets  (\href{https://www.nature.com/articles/s41567-022-01628-6}{Nature Physics volume 18, pages 669–677 (2022)}), suggesting that these may not be one-dimensional objects as speculated. Our work demonstrates the presence of localized defects at the boson peak frequency and provides a comprehensive characterization of their vibrational and geometric properties, resolving the tension between the concepts of quasi-localized quadrupolar defects and stringlets.
\end{abstract}
\maketitle

Thermal and vibrational properties of amorphous solids display universal characteristics that are very different from their crystalline counterparts \cite{doi:10.1142/q0371}. 
In particular, their vibrational density of states (vDOS) $D(\omega)$ presents a universal excess anomaly over the Debye prediction $\textcolor{black}{D(\omega) \propto \omega^{d-1}}$, with $\textcolor{black}{d}$ the number of spatial dimensions and $\omega$ the frequency, known as the ``boson peak'' (BP). 
The microscopic origin of the BP remains one of the most controversial topics in condensed matter physics.

Some theories assume that BP arises from additional nonphononic excitations associated with quasi-localised modes (QLMs).
The observation that structural glasses present nonphononic excitations is old \cite{ROSENSTOCK1962659}. 
This idea was inspired (see \textit{e.g.} \cite{SCHOBER2011501,laird1991localized}) by the concept of resonant modes first formulated in the context of defective crystals \cite{maradudin1963theory} and it is a recurrent theme within the soft potential model of glassy anomalies \cite{doi:10.1080/00018738900101162,PhysRevB.53.11469,PhysRevB.67.094203,PhysRevB.76.064206}. 
These nonphononic excitations have been widely reported in the low-frequency spectrum of amorphous solids and even disordered crystals \cite{PhysRevLett.129.095501}, where they contribute to the density of states with a faster than Debye term, $D(\omega) \propto \omega^p$ with $\textcolor{black}{p>(d-1)}$\cite{PhysRevLett.117.035501,PhysRevE.98.060901,wang2019low,massa2021open,kumar2021density}. 
They have a quadrupolar nature and are characterized by a core made of few atoms accompanied by long-range displacements with power-law decay \cite{10.1063/5.0069477}.
It has been argued that similar defects
%QLMs 
also exist at \textcolor{black}{higher frequencies} and cause the boson peak
~\cite{schober1996low,PhysRevB.67.094203,PhysRevB.76.064206,myPRL2,lerner2023boson,Moriel2024}.
However, this scenario is difficult to prove, as these defects would strongly hybridize with extended phononic modes around the BP frequency. {In particular, methods based on localization properties, such as the participation ratio, are not effective at these frequencies.}
{ If defect exists at high frequencies, they do not appear as QLM, rendering their identification~\cite{PhysRevB.53.11469,PhysRevLett.125.085502} and the study of their properties problematic.}
%they do not appear as QLM, do not appear as modes with a 
%, rendering their identification~\cite{PhysRevB.53.11469,PhysRevLett.125.085502} and the study of their properties problematic.
Henceforth, the existence of localised defects at the boson peak is questionable, motivating the development of alternative theories of the BP \cite{doi:10.1073/pnas.252786999,GiorgioParisi_2003,PhysRevLett.86.1255,PhysRevLett.106.225501,PhysRevLett.122.145501,Marruzzo2013,tanakaNatPhys}.

Recent works have explored the possibility of disentangling 
%QLM
localised defects and extended modes at the boson peak frequency. 
In one approach, deformations with seemingly quadrupolar structures were visually isolated from vibrational modes at the boson peak frequency to assess whether they could serve as seeds for energy-minimal localized deformations~\cite{lerner2023boson,Moriel2024}, which are identified with QLMs. 
However, local quadrupolar deformations are not generally apparent in the hybridized displacement field, as demonstrated in Appendix A, and defects might, in principle, induce diverse localized deformations.
Alternatively, one could examine how modes near the boson peak contribute to the amplitude of particle vibrational motion~\cite{SCHOBER2011501}.
%In contrast, Refs.~\cite{tanakaNatPhys,tanakaPRR} examined how modes near the boson peak contribute to the amplitude of particle vibrational motion, building on earlier works~\cite{SCHOBER2011501}. 
Refs.~\cite{tanakaNatPhys,tanakaPRR} 
%found 
\textcolor{black}{have suggested that}
particles receiving significant contributions from these modes tend to organize along string-like clusters or stringlets { with one-dimensional nature and lacking quadrupolar characteristics}.
The interaction between stringlets and phonons induces a flat boson peak mode~\cite{10.1063/5.0210057,jiang2024phonons}, observed in both simulations~\cite{tanakaNatPhys} and experiments~\cite{tomterud2023observation} supports the hypothesis that stringlets, rather than quasi-localized quadrupolar modes, could be the microscopic origin of the boson peak, even in heated crystals~\cite{10.1063/5.0197386}. 
It remains unclear whether localized defects are responsible for the boson peak in glasses and whether these defects are quasi-localized quadrupolar modes or string-like excitations. 
Thus, the microscopic origin of the boson peak remains unresolved.

In this work, we resolve this tension by introducing a simple and efficient method based on cage-relative \textcolor{black}{(CR)} displacement to disentangle extended phonons from quasi-localized defects. 
We demonstrate that the modes at the boson peak frequency are a superposition of phonons and anisotropic, compact, two-dimensional localized defects, exhibiting pure shear oscillatory deformations in two and three spatial dimensions.
We show that the size of these defects correlates with the length scales associated with the boson peak, providing strong support for previous suggestions that the boson peak originates from quadrupolar quasi-localized excitations~\cite{myPRL2,lerner2023boson}. 
Additionally, we identify the so-called stringlets~\cite{tanakaNatPhys}, revealing that, contrary to earlier speculations, they are not string-like objects but compact ones that align with the localized defects we have uncovered.
Our work unifies different theoretical perspectives on the boson peak in glasses, resolving the long-standing debate and potentially establishing a consensus on its microscopic origin.

We numerically investigate the vibrational properties of two- and three-dimensional systems of polydisperse particles interacting via a Lennard-Jones-like potential $U(r_{ij},x_c)$ which vanishes with its first two derivatives at $x_c \sigma$, $\sigma$ being the average particle diameter of the interacting particles~\cite{dauchotPotential,chattoraj2020role}. 
Previous works have shown that $x_c$, which controls the extension of the attractive well, influences the system's relaxation dynamics~\cite{chattoraj2020role} and the mechanical response~\cite{dauchotPotential, gonzalez2020mechanical, gonzalez2020mechanical2, Zheng2021}. 
%We simulate $N=32k$ particle systems in square or cubic simulation boxes with periodic boundary conditions at fixed interparticle spacing $a_0 = \rho^{-1/D}\simeq 0.977$. 
{ We simulate systems with $N=32k$ particles - which are enough for finite size systems to be negligible at the boson peak frequency \cite{Shintani2008,tanakaPRR,lerner2023boson} - in square or cubic simulation boxes with periodic boundary conditions at fixed interparticle spacing $a_0 = \rho^{-1/D}\simeq 0.977$. }
We generate amorphous solid configurations by minimizing, via conjugate gradient, the energy of systems in thermal equilibrium in the NVT ensemble at $T=4.0\epsilon$, above the glass transition temperature for the considered $x_c$ values~\cite{chattoraj2020role}.
All data presented below are averaged over $200$ configurations at each $x_c$.
We provide details on the model, the numerical procedures, and the $x_c$ dependence of the vibrational model
in Appendix B.

{\it Exposing hybridized defects -- }
%Due to the interaction with phonons, localised vibrations cannot be directly visualized as harmonic vibrations when their frequency is inside a phonon band.
%Instead, the phonons-defect hybridization leads to extended harmonic modes combining the two.
%Devising strategies to disentangle this hybridization and expose the localised defects is a longstanding challenge.
%Two approaches have been introduced so far: one relies on the assumption that the localised defects appear in regions where the local vibrational kinetic energy is high ~\cite{SCHOBER2011501,tanakaNatPhys,tanakaPRR}, and the other on the hypothesis that these defects have a quadrupolar structure visible in the hybridized field~\cite{lerner2023boson, Moriel2024}.
We propose a simple and practical approach to disentangle localised defects and extended plane waves based on the hypothesis that plane waves at the BP frequency are locally affine while localised defects are not. 
%Diverse approaches could be employed to identify regions of local non-affine behaviour \cite{falk1998dynamics}.
\textcolor{black}{Diverse approaches could be employed to decompose a vibration into a phononic and a non-phononic component
~\cite{caroli2020key,shimada2018anomalous}, or more generally, to identify regions of local non-affine behaviour \cite{falk1998dynamics,Flenner2025}.}
Here, we rely on the cage-relative (CR) displacement field to separate and remove the plane-wave contribution from the eigenvectors.
\textcolor{black}{This approach has been previously used to filter out the effect of low-frequency long-wavelength plane waves on the low-temperature liquid dynamics~\cite{vivek2017long,illing2017mermin,Li2019b}, never to isolate defects.}
%Here, we rely on the cage-relative displacement field, which has been previously used to filter out the effect of long-wavelength plane waves~\cite{vivek2017long,illing2017mermin,Li2019b}, to separate and remove the plane-wave contribution from the eigenvectors.
This approach involves analyzing particles' displacements relative to their immediate surroundings or the `cage' formed by neighbouring particles.
Specifically, given a displacement field ${\bm u}$, the cage-relative displacement ${\bm u}^{\rm cr}_{i}$ of particle $i$ is 
${\bm u}^{\rm cr}_{i} = {\bm u}_{i} - \frac{1}{N_v} \sum_{j=1}^{N_v} {\bm u}_{j}$,
where the summation extends over the $N_v$ voronoi neighbors of particle $i$.
We use this approach by considering as displacement fields the vibrational modes of the systems, ${\bm e}_{k}$, which we determine through the direct diagonalization of the Hessian matrix.

\begin{figure}
    \centering
    \includegraphics[width=\linewidth]{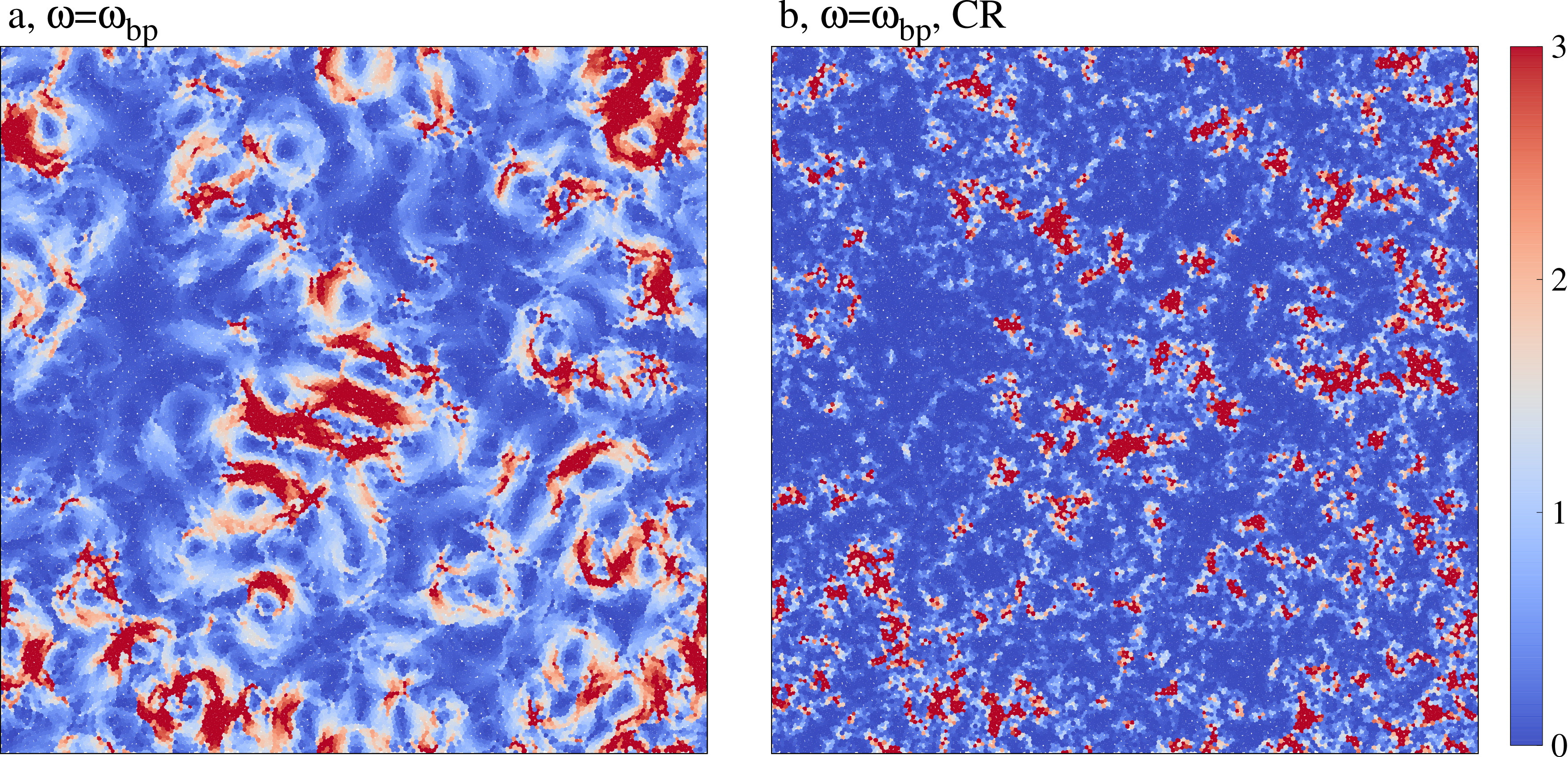}
    \caption{
    \textbf{a} Colour map of the magnitude $|{\bm e}_{k,i}|^2$ of the particle displacement associated with an eigenmode with $\omega_k=\obp$. Swirls characterizing hybridization with plane waves are observed. 
    \textbf{b} The cage-relative displacement, $|{\bm e}_{k,i}^{\rm cr}|^2$ reveals the presence of many vibrational defects not apparent in the standard measure. In these plots, the magnitudes are scaled so that their average value is $1$. 
    }
    \label{fig:modes}
\end{figure}

We validate this approach in Appendix C by demonstrating that the CR measure preserves the localized defects and successfully filters out the plane wave contributions at the boson peak frequency. 
As an example, Fig.~\ref{fig:modes} illustrates a snapshot of a two-dimensional system where each particle is colour-coded based on the magnitude of the displacement $|{\bm e}_{k,i}|^2$ (in panel a), and cage-relative displacement $|{\bm e}_{k,i}^{\rm cr}|^2$ (in panel b), for an eigenmode at the boson peak frequency. 
The cage-relative measure reveals defective regions of large non-affine particle motion. This approach works equally well in three-dimensional systems, as shown in the inset of Fig.~\ref{fig:bp}(c).

We consider as defective particles those with an $|{\bm e}^{~\rm cr}_{k,i}|^2$ value in the top 5\%, unless otherwise specified, and determine the vibrational defects via a cluster analysis of these particles, considering two particles as belonging to the same cluster if voronoi neighbours.
\textcolor{black}{In the following, we discuss results obtained via the analysis of a single mode at the BP frequency, for each system. We have checked that analogous results occur for close frequencies $|\omega-\omega_{\rm bp}| < 0.1 \omega_{\rm bp}$, but not at higher ones.}

%Within a soft potential model perspective, we expose defects coupled among them and with the surrounding medium, not bare defects~\cite{doi:10.1080/00018738900101162, PhysRevB.53.11469, PhysRevB.67.094203, PhysRevB.76.064206}. 

\begin{figure}[!t]
\centering
\includegraphics[width=\linewidth]{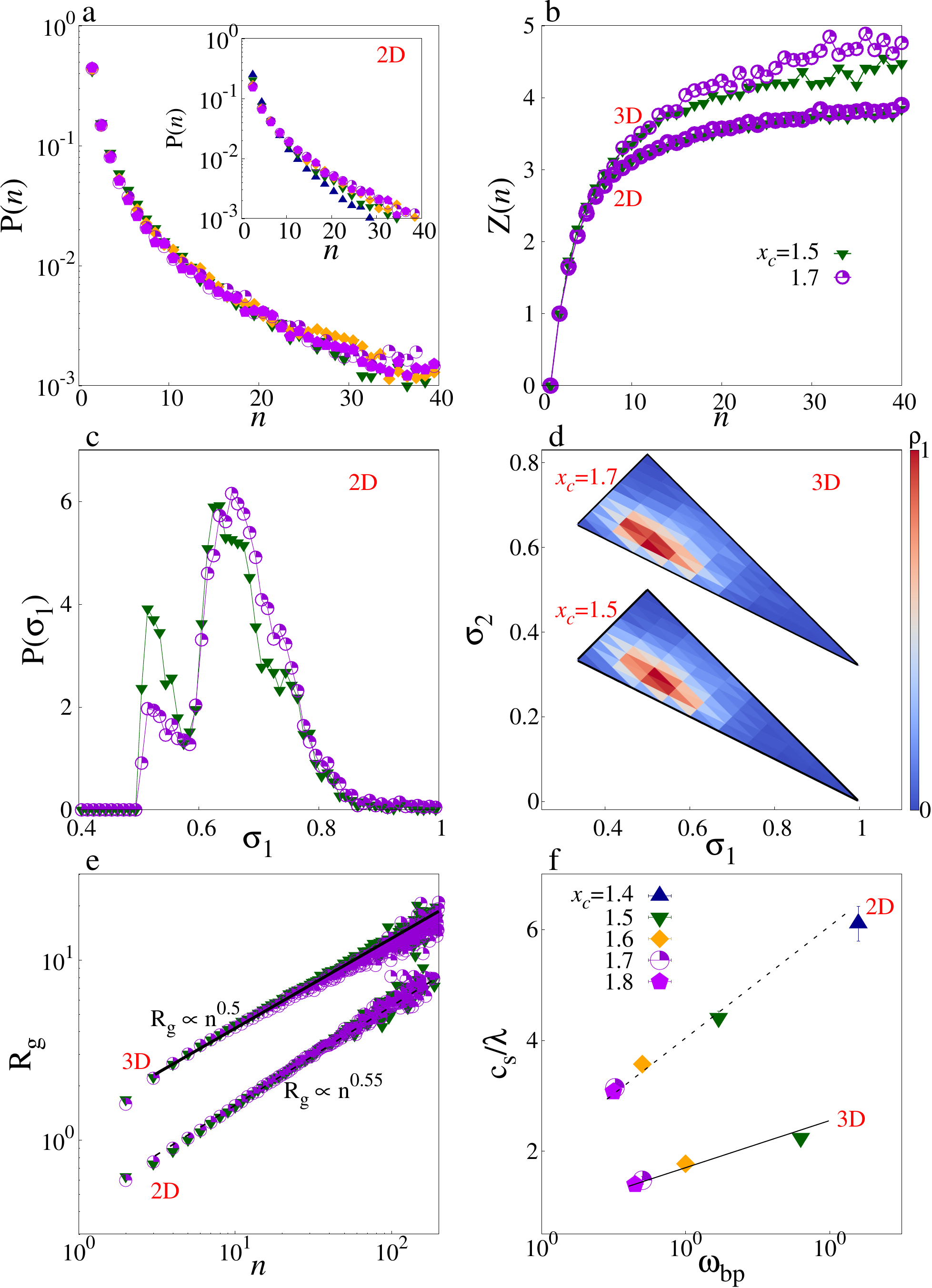}
\caption{
{\bf a} Probability distribution of the number of particles belonging to each defect in 3D and 2D (inset).
{\bf b} Average coordination number of the particles in clusters of size $n$.
{\bf c} We normalise ($\sum \sigma_i = 1$) and order ($\sigma_i > \sigma_{i+1}$) the standard deviations of the particles' positions along the principal axis of each `defect', considering defects with at least $n=3$ particles. 
In two dimensions, the probability distribution of $P(\sigma_1)$ peaks at intermediate values and drops at large ones, indicating that the defects are not one-dimensional ($\sigma_1 = 1$), but two-dimensional objects, albeit not precisely circular $(\sigma_1=1/2)$.
{\bf d} In three dimensions, the probability distribution $P(\sigma_1,\sigma_2)$ clarifies that clusters are not one-dimensional ($\sigma_1 = 1$) or two-dimensional $(\sigma_1+\sigma_2 = 1)$.
We thus assume they are three-dimensional, albeit non-spherical $(\sigma_1 = \sigma_2 = 1/3)$.
Data for $x_c = 1.7$ are shifted vertically for clarity.
{\bf e} Radius of gyration, $R_g$, as a function of cluster size $n$. 3D data are shifted by a factor $3$ for clarity.
{\bf f} The characteristic length scale $\lambda$ of defects governs the boson peak frequency in two and three dimensions.
In all panels, symbols identify the values of $x_c$.
}
\label{fig:geometry}
\end{figure}

{\it Geometrical properties -- }
Fig.~\ref{fig:geometry}(a) shows that the probability distribution of the number of particles in a cluster $P(n)$ decays abruptly at small $n$ and roughly exponentially at larger $n$, similar to the distribution of the number of particles participating in the stringlets \cite{10.1063/5.0039162,10.1063/1.4878502,tanakaNatPhys}. 
On increasing $x_c$, the average cluster size varies from $\langle n \rangle= 16$ to $26$, in 3D, and $\langle n \rangle= 9$ to $13$, in 2D. 
The average coordination number $\textcolor{black}{Z(n)}$ grows with $n$ due to the competition between bulk and surface particles, as in Fig.~\ref{fig:geometry}(b), and it is larger than two if not for minute clusters, indicating that the clusters are not string-like.
%Indeed, the radius of gyration scales with the cluster size as $R_g\propto n^q$, with $q\simeq0.55$ in 2D and $q \simeq0.5$ in 3D, if not for $n=2$ where clusters have to be linear.
%suggesting the clusters may be two-dimensional objects.

To further assess a cluster's geometrical properties, we perform a principal component analysis of the covariance matrix $C_{\alpha\beta}=\langle r_\alpha r_\beta\rangle-\langle r_\alpha\rangle \langle r_\beta\rangle$ of the positions of its particles.
We use the eigenvalues $\lambda_i$ of $C_{\alpha\beta}$, which measure the variances of the particle positions along the principal directions, to define normalized standard deviations $\sigma_i = \sqrt{\lambda_i} / \sum_i \sqrt{\lambda_i}$ that obey the geometrical constraint $\sum_i \sigma_i = 1$, and order them such that $\sigma_{i} > \sigma_{i+1}$. 
In two dimensions, $\sigma_1 = 1$ for linear (maximally anisotropic, one-dimensional) clusters and $\sigma_1 = \frac{1}{2}$ for circular (isotropic) clusters. 
We find that the distribution $P(\sigma_1)$ peaks at $\sigma_1 \simeq 0.65$ and vanishes for $\sigma_1 > 0.8$, as shown in Fig.~\ref{fig:geometry}(c).
In three dimensions, for linear (1D), flat (2D), and spherical (isotropic 3D) clusters, $\sigma_1 = 1$, $\sigma_1 + \sigma_2 = 1$, and $\sigma_1 = \sigma_2 = \sigma_3 = \frac{1}{3}$, respectively. 
Fig.~\ref{fig:geometry}(d) presents density maps of the $P(\sigma_1, \sigma_2)$ probability distribution at $x_c = 1.5$ and $x_c = 1.7$ (shifted vertically) and suggests that the localised defects are three-dimensional compact (panel b) clusters with a high degree of anisotropy. 
{ 
Due to this anisotropy, the radius of gyration scales as $R_g\propto n^q$, with $q\simeq0.55$ in 2D and $q \simeq 0.5$ in 3D, if not for large clusters, as shown in Fig.~\ref{fig:geometry}(e).
}
%suggesting the clusters may be two-dimensional objects.
The distribution of the clusters' radius of gyration is exponentially distributed (see Appendix D), allowing us to associate a typical length scale $\lambda$ to each $x_c$. Fig.~\ref{fig:geometry}(f) shows that $\obp \propto c_s/\lambda$, in both two and three dimensions, proving these defects' relevance to the boson peak phenomenology. 

%\textcolor{black}{These defects obtained at $\obp$, suggests a relationship between these defects and the frequency scale of the boson peak. So, we test more rigorously whether defects regulate the boson peak by evaluating their characteristic length scale $\lambda=n^q$ from their exponential distributions (see Fig.S5). We show that $\obp \propto c_s/\lambda$ (panel f), in both two and three dimensions, again demonstrating these defects regulate the boson peak.}

\begin{figure}[!t]
    \centering
    \includegraphics[width=\linewidth]{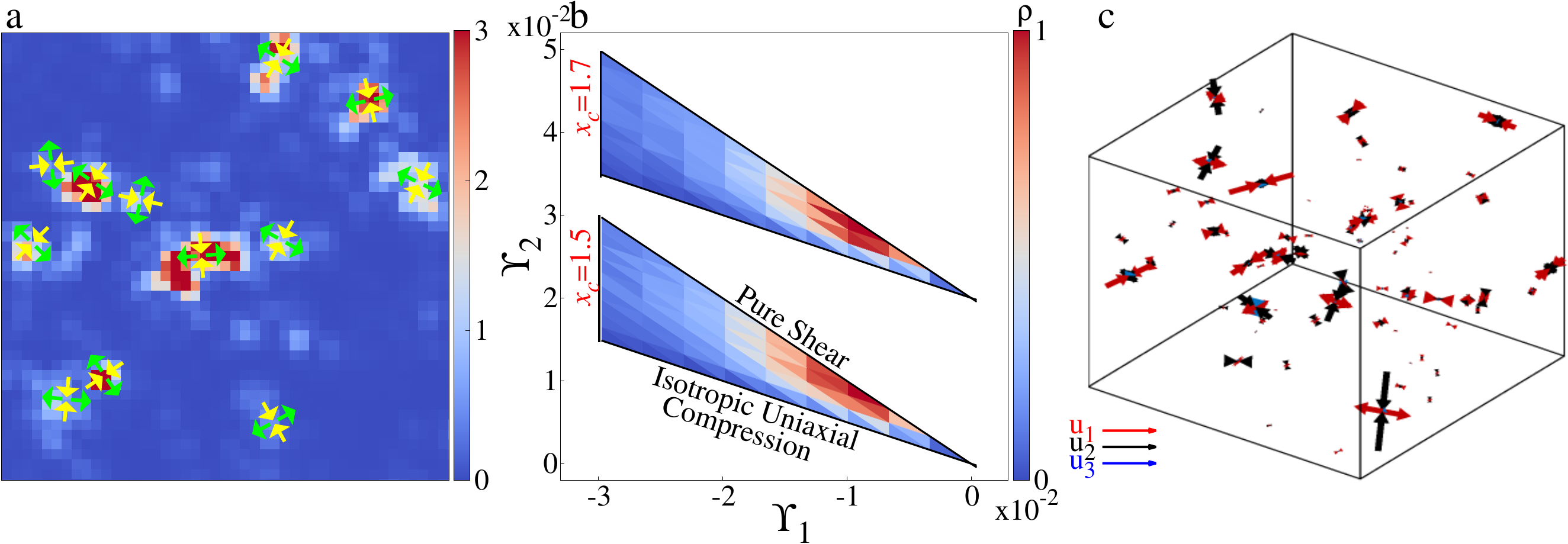}
    \caption{
    %{\bf a} Pearson correlation coefficient between coarse-grained cage-relative displacement and quadrupolar charge.
    {\bf 1} Heat map of the magnitude of the coarse-grained cage-relative displacement, normalised by its average value, with superimposed quadrupolar displacements of the defects. The figure refers to $x_c =1.7$ and depicts 1/12th of the system's area.
    {\bf b} Density map of the two largest (in modulus) eigenvalues of the deviatoric strain associated with the defects, normalised to its maximum value. The density is higher close to the $\Upsilon_2=-\Upsilon_1$ line, implying that the quadrupoles correspond to pure shear deformations.
    {\bf c} We represent the quadrupolar deformations by attaching to each defect's centre of mass the displacements along the principal axis $\pm \Upsilon_1{\bm u}_1$ (blue), $\pm \Upsilon_2{\bm u}_2$ (green) and $\pm \Upsilon_3{\bm u}_3$ (red). Since quadrupoles mostly correspond to pure-shear deformations, ${\bf u}_3$ is barely visible.
    }
    \label{fig:quadrupolar}
\end{figure}
{\it Vibrational properties -- }
\textcolor{black}{
To investigate the vibrational properties of these clusters, we first evaluate the strain tensors coarse-grained over a small length scale. 
Since the strain is insensitive to local uniform translations, as the cage-relative measure, the cage-relative clusters occur in regions of high local strain.
This allows us to evaluate a defect strain tensor $\mathcal{E} =\mathcal{E}_{\rm vol}+\mathcal{E}_{\rm dev}$, as detailed in Appendix E.
We find that these tensors are dominated by their deviatoric component, indicating that defects perform volume-preserving deformations, as $R=\frac{\sqrt{Q}}{|\rm{Tr}(\mathcal{E})|} \gg 1$. 
Here, $Q=1/2\sum \Upsilon_k^2$ is the quadrupolar charge and $\Upsilon_k$ are the eigenvalues of the traceless $\mathcal{E}_{\rm dev}$, and $|\rm{Tr}(\mathcal{E})|$ quantifies the degree of volumetric deformations.
In the following, we assume $\Upsilon_1 < 0$, order the eigenvalues so that $|\Upsilon_i|>|\Upsilon_{i+1}|$, and indicate their associated eigenvectors with ${\bf U}_k$.
} A cluster extends into the directions with $\Upsilon_k > 0$ while it shrinks in the others for half a period of oscillation, with particles reversing their direction of motion during the remaining half period. 
%In the following, we assume $\Upsilon_1 < 0$ and order the eigenvalues so that $|\Upsilon_i|>|\Upsilon_{i+1}|$.

The traceless condition implies $\Upsilon_1 = -\Upsilon_2$ in two dimensions.
\textcolor{black}{Figure~\ref{fig:quadrupolar}(a)} illustrates the quadrupoles associated with each cluster on a heat map of the magnitude of the coarse-grained cage-relative displacement, finding expected correlations. % in light of panel (a).
The figure illustrates a small fraction of the system and represents each quadrupole with four arrows with a length proportional to $\pm \Upsilon_i {\bm U}_i$. See Fig.~\ref{fig:geometry2d} in Appendix A for a snapshot of the whole system and illustrations of the underlying displacement field.

{ In three dimensions, different deformation modes satisfy the traceless condition $\sum \Upsilon_i = 0$. 
We expose the dominant deformation mode by analyzing the probability distribution $P(\Upsilon_1, \Upsilon_2)$. }
%In three dimensions, we examine the relevance of various deformation modes that satisfy the traceless condition $\sum \Upsilon_i = 0$ by analyzing the probability distribution $P(\Upsilon_1, \Upsilon_2)$. 
In the $\Upsilon_1$-$\Upsilon_2$ plane, this distribution is supported within a region bounded by the line $\Upsilon_1 + \Upsilon_2 = 0$, which represents pure shear deformation, and the line $\Upsilon_1 + 2\Upsilon_2 = 0$, which corresponds to volume-preserving isotropic ($\Upsilon_3 = \Upsilon_2 = -\Upsilon_1/2$) uniaxial compressions. 
\textcolor{black}{Figure~\ref{fig:quadrupolar}(b)} shows that quadrupolar deformations are predominantly pure shear, as $P(\Upsilon_1, \Upsilon_2)$ peaks near the $\Upsilon_1 + \Upsilon_2 = 0$ line for different values of $x_c$. 
This result is further supported by direct visualization of the quadrupoles (\textcolor{black}{Figure~\ref{fig:quadrupolar}(c)}), depicted as collections of 6 arrows with lengths proportional to $\pm \Upsilon_i {\bm U}_i$. 
Since $|\Upsilon_3| \ll |\Upsilon_2|$, the blue arrows corresponding to displacements in the ${\bm U_3}$ direction are not visible, making the quadrupoles appear as a two-dimensional cross.
{ Overall, these investigations show that modes at the boson peak comprise plane waves hybridized with two-dimensional defects that perform pure shear deformations, in both two- and three-spatial dimensions.}

\begin{figure}[!t]
\begin{center}        
    \includegraphics[width=\linewidth]{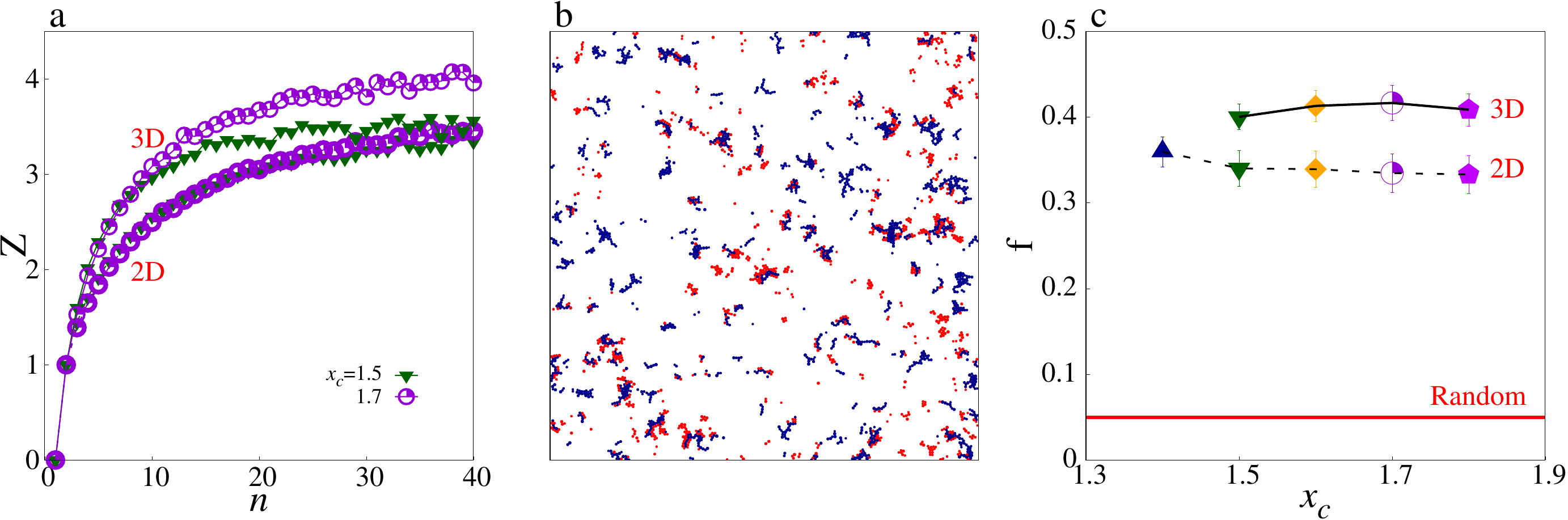}
\end{center}
\caption{
{\bf a} The coordination number of vibrability clusters increases with $n$, following a trend similar to that of clusters formed based on cage-relative displacement (see Fig.3b).
{\bf b} Particles exhibiting high cage-relative displacement at the boson peak (red) and high vibrability (blue), both in the top 5\%, for $x_c = 1.7$. 
Strong correlations between the two are observed. 
{\bf c} The fraction of particles displaying both large cage-relative displacement and high vibrability significantly exceeds the expectation under random conditions. }
\label{fig:peace}
\end{figure}
{\it The connection with `stringlets' -- }
%At this point, one might then wonder how our findings are compatible with the recent claims by Hu and Tanaka \cite{tanakaNatPhys}, that on the contrary identify the microscopic origin of the BP in one-dimensional string-like dynamical objects. 
%Hu and Tanaka \cite{tanakaNatPhys} have identified the particles with a large vibrational amplitude, or equivalently, a high `vibrability' $\Psi_i = \sum_k \frac{1}{\omega_k^2} \left| {\bf e}_{k,i} \right|^2$, where the sum extends over all modes with frequencies $\omega_k$ close to the boson peak frequency.
At this point, one might then wonder how our findings are compatible with the recent claims by Hu and Tanaka \cite{tanakaNatPhys}, who suggested that the BP is induced by one-dimensional clusters, or `stringlets'. 
These are clusters of particles with a large vibrational amplitude, or equivalently, a high `vibrability' $\Psi_i = \sum_k \frac{1}{\omega_k^2} \left| {\bf e}_{k,i} \right|^2$, where the sum extends over all modes with frequencies $\omega_k$ close to the boson peak frequency.
% $|\omega_{\text{bp}} - \omega_k| < \Delta \omega$.
%{ They suggested that these particles organize in one-dimensional clusters, or `stringlets', which induce the boson peak.}
%They have found these particles organize in linear-looking clusters, or `stringlets', and suggested these stinglets induce the boson peak.
We have repeated their analysis in our simulation model. 
Figure~\ref{fig:peace}(a) shows that the average coordination number of the particles participating in the resulting cluster is large, indicating that these are compact and extended objects, rather than one-dimensional, very similar to the localised defects investigated in Fig.~\ref{fig:geometry}(b).
In Figure~\ref{fig:peace}(b), we observe a high spatial correlation between particles with vibrability in the top 5\% (blue) alongside those with cage-relative displacements in the top 5\% (red), indicating a high correlation between vibrational motion and cage-relative displacement at the BP frequency. 
These correlations are further confirmed by measuring the fraction $f$ of particles with high vibrability and cage-relative displacement, which is much larger than expected in the absence of correlations, as we illustrate in Figure~\ref{fig:peace}(c). Overall, our analysis provides strong evidence that quadrupolar localised defects and high-vibrability clusters are \textit{two sides of the same coin}. 
In our systems, however, these clusters are not string-like, and their motion is quadrupolar rather than one-dimensional.

\color{black}

{\it Conclusions -- }
%We have identified localised defects hybridizing with plane waves at the boson peak frequency through a filtering technique based on the physical insight that localized vibrations are generally less affine than extended ones. 
%This technique allowed us to isolate the vibrational defects and study their geometrical and vibrational properties.
%The vibrational defects are spatially anisotropic and perform a quadrupolar oscillatory motion.
%Specifically, the defects undergo pure shear oscillatory deformations, extending along one axis while contracting along a perpendicular one in both two and three spatial dimensions. 
These findings support earlier results~\cite{myPRL2,lerner2023boson} suggesting a relation between the boson peak and quasi-localised vibrational modes in a soft-potential model perspective~\cite{doi:10.1080/00018738900101162, PhysRevB.53.11469, PhysRevB.67.094203, PhysRevB.76.064206}.
Our approach has the merit of allowing for a straightforward identification of the hybridized defects in 2D, when these are not visible in the displacement field, and in 3D.
The connection between our defects and QLMs deserves further investigation.
In particular, while our defects exclusively undergo pure shear deformations, Ref.~\cite{Moriel2020} reported a low-frequency QLM undergoing simple shear deformations where extension occurs in one direction and contraction in the orthogonal ones. 
\textcolor{black}{
In addition, we find that the displacement is predominantly two-dimensional, even in three dimensions, and hence, it does not align with that of the low-frequency localized modes~\cite{kapteijns2018universal}.}

%At this point, one might wonder how our findings are compatible with the recent claims by Hu and Tanaka~\cite{tanakaNatPhys}, which instead attribute the microscopic origin of the BP to one-dimensional string-like dynamical objects. To investigate this, we perform a stringlet analysis following their approach, as detailed in Appendix E. Our results reveal that the localized defects identified in our work correspond closely to those found in their study, though they are compact and quadrupolar rather than string-like.

Along the lines of \cite{myPRL2}, it would be interesting to understand whether the same defective structures could also be rationalized using shear modulus fluctuations and whether our findings are compatible with another successful theory of the BP -- heterogeneous elasticity theory \cite{doi:10.1142/9781800612587_0009}. 
Moreover, it would be fruitful to investigate the relation between the defects discussed in our manuscript and (i) the geometric charges responsible for plastic screening and anomalous elastic response~\cite{PhysRevE.104.024904}, (ii) the topological vortex-like defects recently connected to plastic soft spots~\cite{Wu2023, Baggioli2023}.

{\it Acknowledgements-} MPC acknowledges support by the Singapore Ministry of Education through grants MOE-T2EP50221-0016 and RG152/23. CJ and MB acknowledge the support of the Shanghai Municipal Science and Technology Major Project (Grant No.2019SHZDZX01). MB acknowledges the support of the sponsorship from the Yangyang Development Fund.

%\bibliography{Maintext}
%% paste the bib file here when done, and comment the previous line


\newpage

\appendix
\section{Extracting defects from a disordered vibrational mode} 
\begin{figure}[h]
\begin{center}        
    \includegraphics[width=\linewidth]{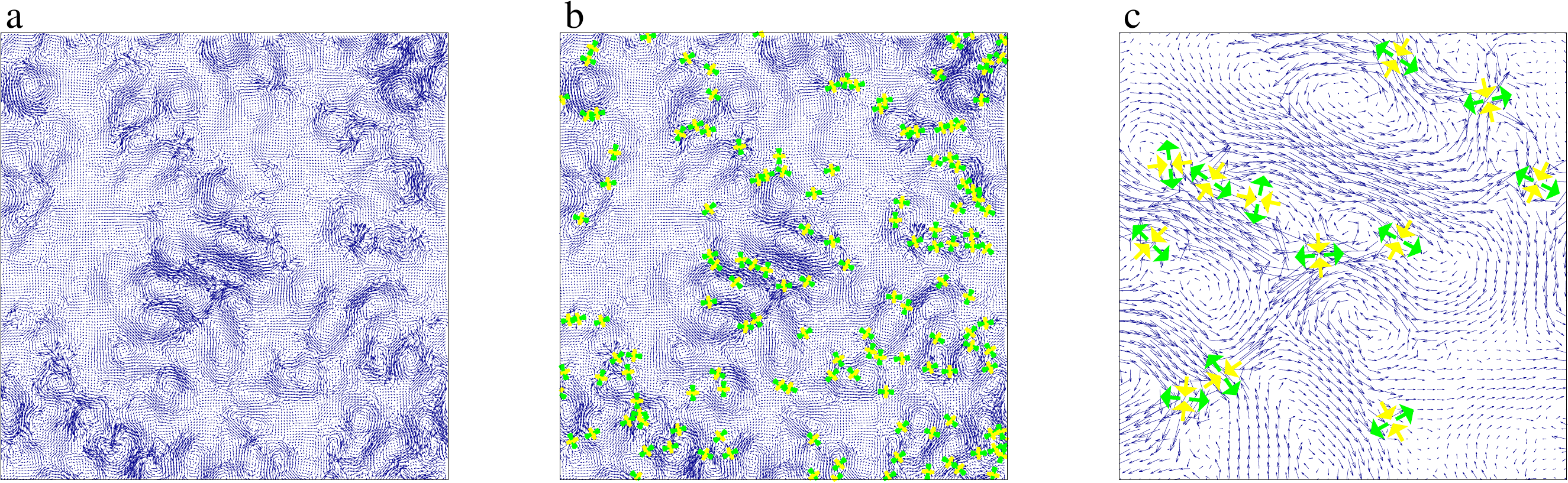}
\end{center}
\caption{
{\bf a} Eigenvector at the boson peak frequency for a two-dimensional system with $x_c = 1.7$. It is arduous to spot in similar modes isolated quadrupolar-looking defects~\cite{lerner2023boson,Moriel2024}.
{\bf b} Defects identified by our approach.
{\bf c} A zoom on a region corresponding to 1/12th of the system's area. }
\label{fig:geometry2d}
\end{figure}

Fig.~\ref{fig:geometry2d}a presents an eigenmode at the boson peak frequency, characterized by swirls that indicate hybridization with plane waves. 
This hybridization prevents visually isolating quadrupolar-like defects as recently proposed~\cite{lerner2023boson,Moriel2024}. 
Fig.~\ref{fig:geometry2d}b displays the defects identified by our proposed filtering technique, while panel c provides a magnified view of a region covering 1/12th of the system's area, offering a clearer visualization of the quadrupolar-like localized defects.

\section{Numerical Model} 
We consider two- and three-dimensional systems of particles of diameter $\sigma_i$, drawn from a uniform random distribution in the range [0.8:1.2], that interact via family of LJ-like potentials consisting of a repulsive and an attractive part, $U(r_{ij},x_c)= U_r(r_{ij})+U_a(r_{ij},x_c)$~\cite{dauchotPotential,chattoraj2020role}. 
The repulsive part follows the standard LJ functional form, 
\begin{equation}
U_r(r_{ij})=4\epsilon\left[\left(\frac{\sigma_{ij}}{r_{ij}}\right)^{12}-\left(\frac{\sigma_{ij}}{r_{ij}}\right)^{6}\right],
\label{eq:repPot}
\end{equation}
where $\sigma_{ij}=(\sigma_i+\sigma_j)/2$, but only acts for $r_{ij} \leq r_{ij}^{\rm min}=2^{1/6}\sigma_{ij}$.
The attractive part, which only acts for distances in the range
$2^{1/6}\sigma_{ij} \leq r_{ij} \leq x_c \sigma_{ij}$,
is given by
\begin{equation}
U_a(r_{ij})=\epsilon\left[a_0\left(\frac{\sigma_{ij}}{r_{ij}}\right)^{12}\hspace{-0.4cm}-a_1\left(\frac{\sigma_{ij}}{r_{ij}}\right)^{6}\hspace{-0.2cm}+\sum_{l=0}^3{c_{2l}\left(\frac{r_{ij}}{\sigma_{ij}}\right)^{2l}}\right].
\label{eq:attPot}
\end{equation}
The parameters $a_0,a_1$ and $c_{2l}$ are chosen such that the potential $U(r_{ij})$ and its first two derivatives are continuous at the minimum $r_{ij}^{\rm min}$ and at the cutoff $r_{ij}^{(c)}=x_c \sigma_{ij}$, where the potential also vanishes. 
The parameter $x_c$ sets the extension of the attractive well~\cite{dauchotPotential}, which vanishes at $x_c \sigma$. 
This parameter influences the relaxation dynamics~\cite{chattoraj2020role} and the mechanical response~\cite{dauchotPotential, gonzalez2020mechanical, gonzalez2020mechanical2, Zheng2021}.

\begin{figure}[!!h]
\begin{center}        
\includegraphics[width=\linewidth]{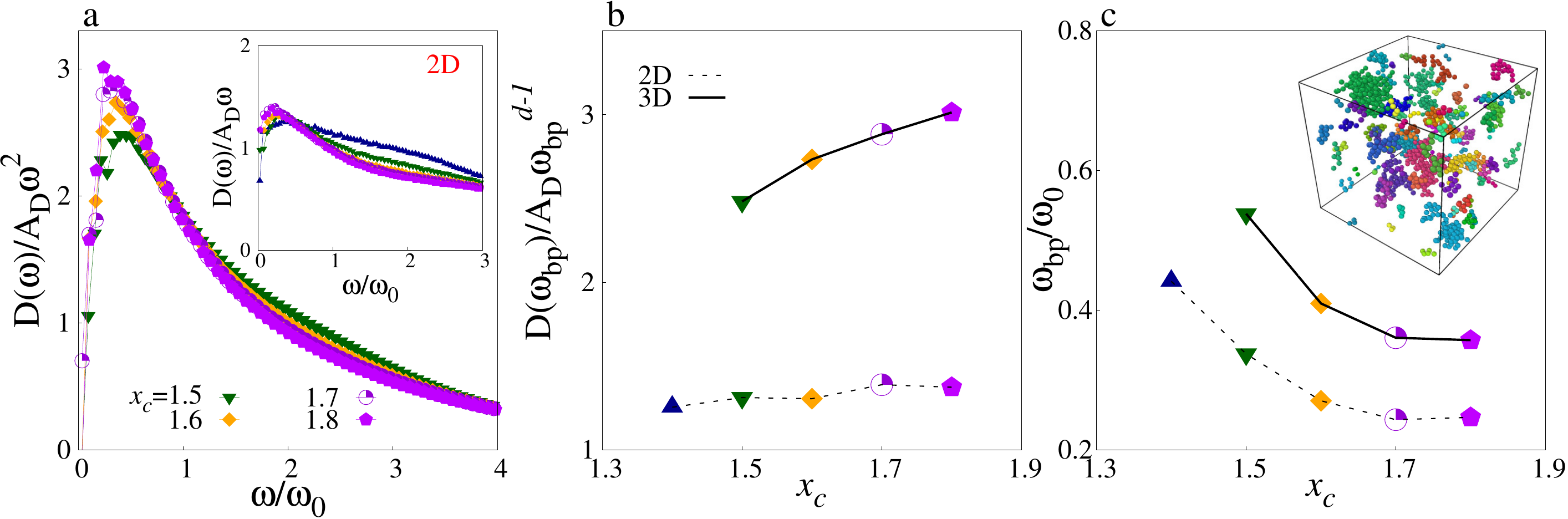}
\end{center}
\caption{
{\bf a} The reduced vDOS in three and (inset) two dimensions for systems differing in the extension $x_c$ of the intermolecular attractive potential well. 
In 3D, the materials lose stability as $x_c$ increases, as the boson peak intensity increases ({\bf b}) while its frequency decreases ({\bf c}). 
In 2D, attraction does not sensibly influence the excess vibrational modes.
Frequencies are expressed in units of the natural frequency $\omega_0 = c_s/a_0$, with $a_0$ the interparticle spacing and $c_s$ the shear wave speed.
The inset of ({\bf c}) illustrates the localised defects that, according to our investigation, induce the boson peak in a three-dimensional system with $x_c = 1.7$.}
    \label{fig:bp}
\end{figure}

\begin{figure}[!!h]
 \centering
 \includegraphics[angle=0,width=0.3\textwidth]{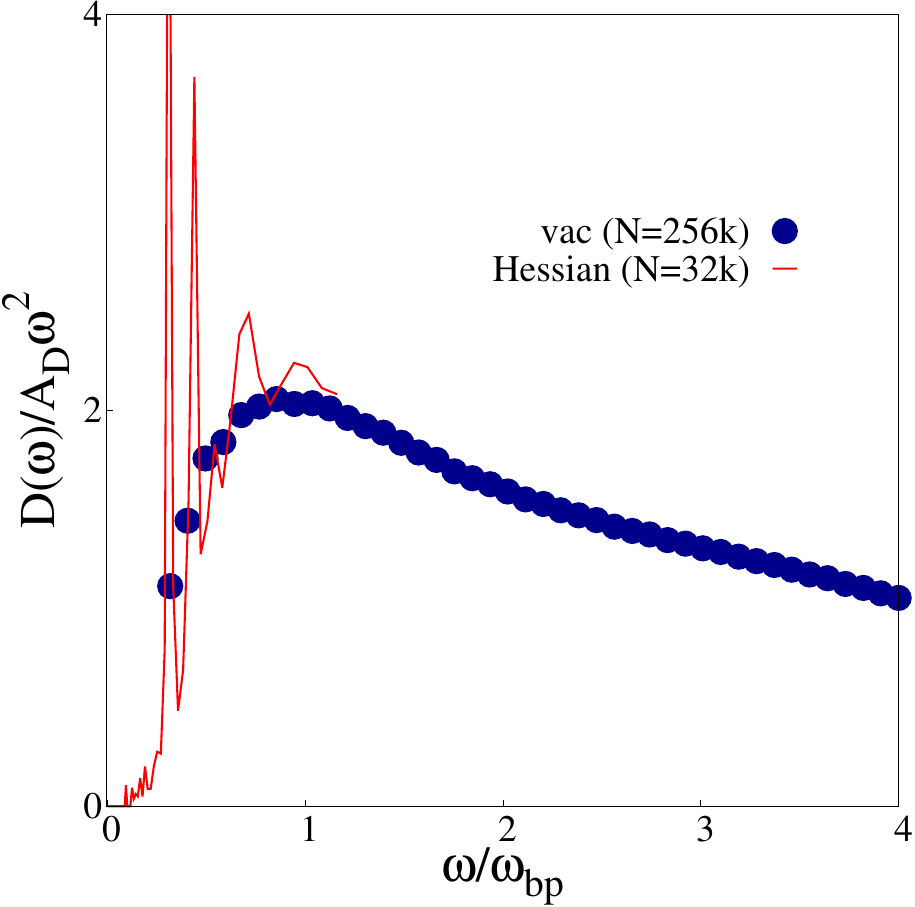}
 \caption{
Density of states normalized by Debye's prediction for $x_c=1.4$ in 3d. In finite systems, at low frequencies, the discrete nature of the spectrum emerges, and hence, $D(\omega)/A_D\omega^2$ approaches zero rather than a constant. At small $x_c$, the discrete nature of the spectrum extends to higher frequencies, making it difficult to identify the boson peak frequency.
\label{fig:band}
}
\end{figure}

We evaluate the vDOS over a broad frequency range by computing the Fourier transform of the velocity autocorrelation function in the linear response regime \textcolor{black}{by linearizing the equation of motion, implying the absence of anharmonic effects.} 
Additionally, we obtain the vDOS by diagonalizing the Hessian matrix, which provides access to eigenvectors, ${\bf e}_k$, and eigenfrequencies, $\omega_k$, in the low-frequency regime, including the boson peak. These methods yield consistent results.
Fig.\ref{fig:bp}(a) shows the reduced vDOS, $\textcolor{black}{D(\omega)/A_D\omega^{d-1}}$, with the Debye prediction, $\textcolor{black}{D_{\rm Debye} = A_D\omega^{d-1}}$, highlighting the BP in two and three spatial dimensions. \textcolor{black}{Here, 
$A_D = d/\omega_D^d$ where $d$ is dimension, $\omega_D^d = \frac{18\pi^2\rho}{c_l^{-3} + 2c_s^{-3}}$ in 3d and $\omega_D^d = \frac{8\pi\rho}{c_l^{-2} + c_s^{-2}}$ in 2d, with the longitudinal and shear velocities given by $c_l = \sqrt{\frac{K + \frac{4}{3}\mu}{\rho}}$ in 3d, $c_l = \sqrt{\frac{K + \mu}{\rho}}$ in 2d and $c_s = \sqrt{\frac{\mu}{\rho}}$ in both 2d and 3d. Here, $K$ is the bulk modulus and $\mu$ is the shear modulus, which we measure by investigating the stress-strain relationships in the linear response regime.
}
In 3D, increasing $x_c$ raises the peak amplitude \textcolor{black}{(see Fig.~\ref{fig:bp}b)} while lowering its frequency, $\obp$ \textcolor{black}{(panel c)}, suggesting that shorter attractive wells stabilize glasses, as previously proposed~\cite{myPRL2,gonzalez2020mechanical,Zheng2021,dauchotPotential}. In 2D, the effect of $x_c$ is weaker,
\textcolor{black}{indicating that the material stability, which depends on $x_c$ in our model, has a dimensionality-dependent influence on the BP. It remains to be ascertained whether these results are universal across different systems.}
For 3D, we analyze $x_c \geq 1.5\sigma$, as smaller values maintain a phonon band structure at high frequencies, complicating $\obp$ identification, as illustrated in Fig.~\ref{fig:band}.
In 2D, we consider $x_c \geq 1.4\sigma$, since for lower values the BP broadens, making $\obp$ difficult to estimate.
 
\section{Cage-relative filtering approach} 
\begin{figure}
    \centering
    \includegraphics[width=0.8\linewidth]{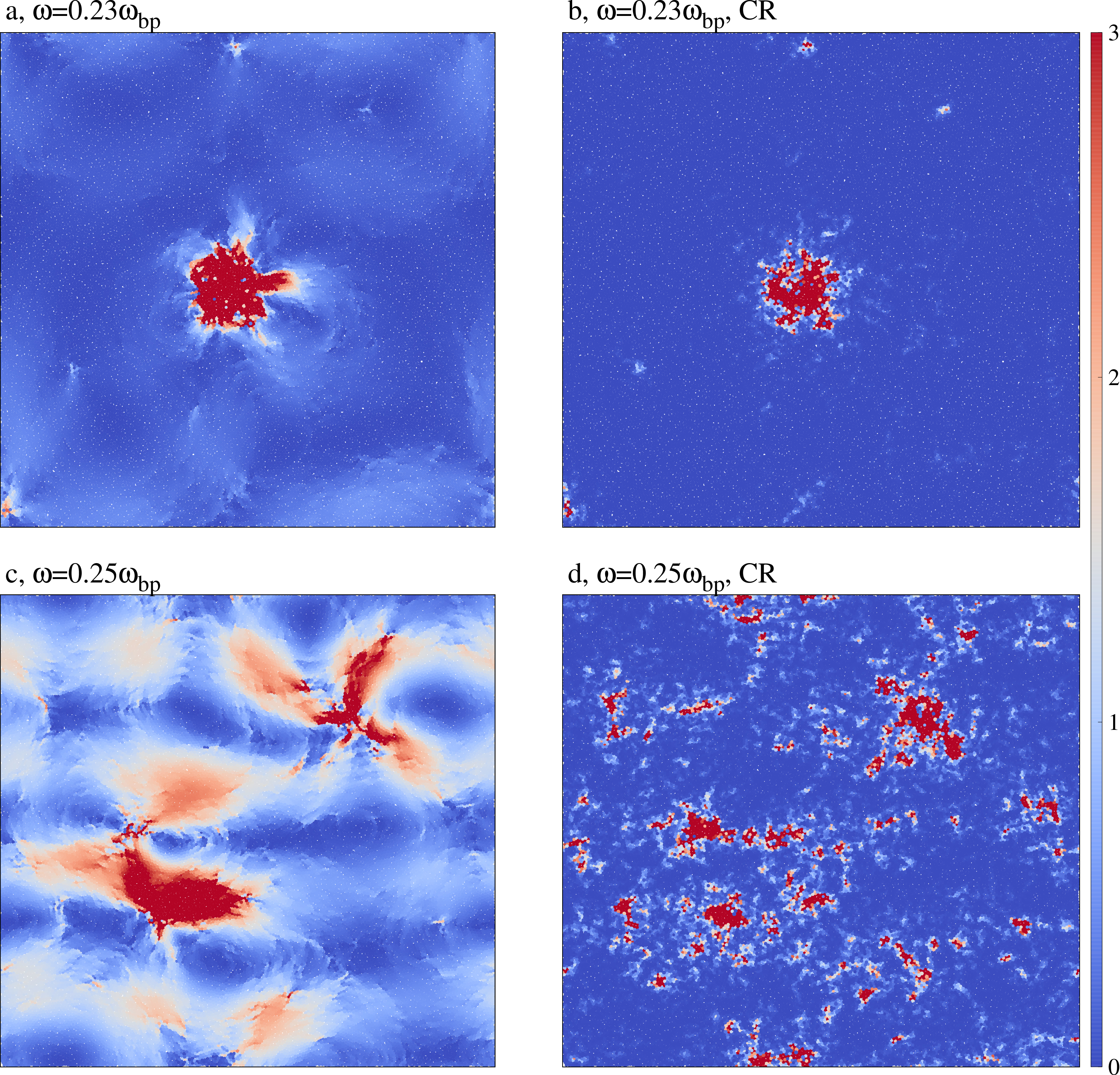}
    \caption{
    \textbf{a} Colour map of the magnitude $|{\bm e}_{k,i}|^2$ of the particle displacement associated with a low-frequency mode with $\omega_k=0.23 \obp$. A localized defect is visible. 
    \textbf{b} The defect remains visible in the colour map of the magnitude of the cage-relative displacement, $|{\bm e}_{k,i}^{\rm cr}|^2$. 
    In all plots, the magnitudes are scaled so that their average value is $1$.
    \textbf{c,d} report analogous results at a similar frequency for a hybridized rather than a quasi-localized mode.
    }
    \label{fig:modesAB}
\end{figure}

To prove that our filtering approach does not filter out localized vibrations, we illustrate in Fig.~\ref{fig:modesAB}(a) a snapshot of a two-dimensional system where each particle is coloured based on its $|{\bm e}_{k,i}|^2$, for a low-frequency localised mode $k$. 
The localised excitation remains visible in panel (b), where particles are coloured according to the magnitude of their cage-relative displacement, $|{\bm e}_{k,i}^{\rm cr}|^2$, proving that our approach preserves localized defects.
\textcolor{black}{Panels (c) and (d) illustrate the same results for a hybridized plane wave.}

\begin{figure}[h]
 \centering
 \includegraphics[angle=0,width=0.30\textwidth]{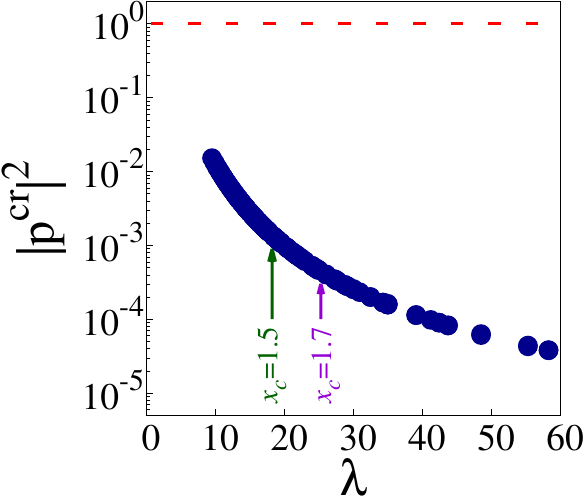}
 \caption{
Squared norm of the cage-relative displacement field $|{\bm p}^{\rm cr}|^2$ associated with phonons ($|{\bm p}|^2=1$) of wavelength $\lambda$, in two dimensions.
The vertical lines mark the wavelength $\lambda_{\rm bp}$ of phonons at the boson peak frequency for the extremal investigated $x_c$ values.
Analogous results hold in 3D.
\label{fig:cr}
}
\end{figure}

To evaluate the cage-relative ability to filter extended modes of norm $|{\bm u}|^2 = 1$, we consider $|{\bm u}^{\rm cr}|^2$.
Filtering occurs if $|{\bm u}^{\rm cr}|^2 \ll 1$.
As extended modes, we consider phonons with various wavelengths.
There is no need to consider the superposition of phonons as the cage-relative `operator' is linear,
$(a{\bm u}+b{\bm v})^{\rm cr} = a{\bm u}^{\rm cr} + b{\bm v}^{\rm cr}$.
%Fig.~\ref{fig:cr} demonstrate that our approach successfully filters out the phonon-induced particle displacements.
Fig.~\ref{fig:cr} illustrates $|{\bm p}^{\rm cr}|^2$ for phonon ${\bf p}$ of wavelength $\lambda$, in two dimensions.
Filtering is more efficient at larger wavelengths, where the induced deformation is more affine.
Importantly, filtering is highly efficient at the wavelength $\lambda_{\rm bp}=\frac{2\pi c_s}{\omega_{\rm bp}}$ of the phonons expected to hybridize at the boson peak, as $|{\bm p}^{\rm cr}|^2$ is in the range $10^{-4}$ to $10^{-3}$, its value depending on the cutoff $x_c$ of the interaction potential.
%\textcolor{black}{After thorough validation of our approach, we compare the magnitude of the displacement and cage-relative displacement for the mode at the boson peak frequency (see main text) that reveals the presence of many vibrational defects not apparent in the standard measure.}

\section{Cluster linear size}
\begin{figure}[!!h]
\centering
\includegraphics[angle=0,width=0.44\textwidth]{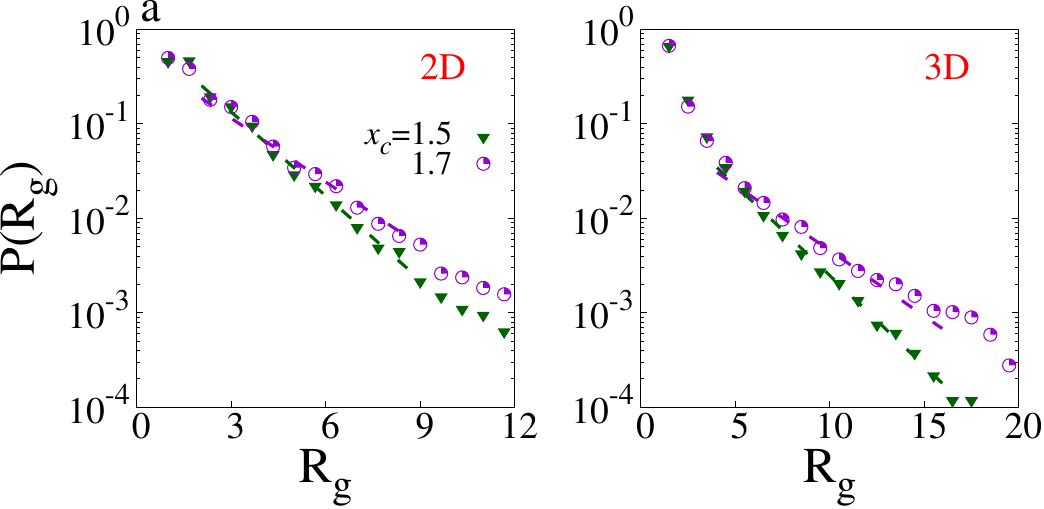}
\caption{
Distribution of the defects' radius of gyration, in 2D and 3D, for diverse values of $x_c$.
\label{fig:RgCR}
}
\end{figure}

Fig.~\ref{fig:RgCR} shows that the radius of gyration of the clusters is exponentially distributed, at large $R_g$, $P(R_g)=ae^{-R_g/\lambda}$, which allow us to associate a typical length scale $\lambda$ to each system.
%We associate with each cluster a length $\lambda=n^q$
%\textcolor{black}{After establishing that the clusters of varying sizes $n$ exhibit characteristics more akin to two-dimensional objects, we show that the linear size of these clusters, defined as $\lambda=n^q$, follows an exponential distribution, $P(\lambda)=ae^{-\lambda/\lambda_0}$ with a characteristic length $\lambda_0(x_c)$ which increases with $x_c$. This result, shown in Fig.~\ref{fig:RgCR} indicates that the cluster size distribution $P(n)$ deviates from a purely exponential form, particularly at small $n$, consistent with previous findings~\cite{jiang2024stringlet}. Specifically, the distribution follows $P(n)=aqn^{q-1}e^{-n^q/\lambda_0}$, which we have verified independently for our systems.}

\section{Cluster deviatoric strain}
We evaluate the local strain tensor 
$
{\bm \epsilon}= \frac{1}{2} \left({\bm \nabla} {\bm{e}} + ({\bm \nabla} \bm{e})^T\right),
$ associated with the displacement field (eigenvector, ${\bm e}$) of the mode at the boson peak frequency, coarse-grained on grids of side-length $w \simeq 1.1$.
We decompose this tensor into its volumetric and deviatoric components:
$
\bm{\varepsilon} = \bm{\varepsilon}_{\rm vol} + \bm{\varepsilon}_{\rm dev}% = m \bm{I} + Q \bm{A}^{ts}
$. 
Here, $\bm{\varepsilon}_{\rm vol}$ represents local volumetric deformations. 
We define a cluster deviatoric strain as $\mathcal{E}_{\rm dev} = \sum \bm{\varepsilon}_{\rm dev}$, where the sum extends over all grid cells containing particles of the considered cluster.

%{\it Appendix F: Phonon bands at small $x_c$ -}
%The boson-peak frequency can only be identified if, at that frequency, the density of states is continuous. The transition to the continuous regime occurs at smaller frequencies (compared to the typical frequency $\omega_0=c/a_0$ with $c$ the sound velocity and $a_0$ the distance between close molecules) in unstable rather than stable glasses - because the former contains more defects that induce a broadening of peaks.

%In our systems, where small $x_c$ corresponds to more stable glasses, at small $x_c$ we observe signatures of a boson peak with superimposed peaks for the $N$ values at which we can diagonalize the vibrational spectrum. 
%This superimposition prevents us from confidently estimating the boson peak frequency. As an example, Fig.~\ref{fig:band} shows the interplay between the discrete spectrum and the boson peak for $x_c=1.4$.

\end{document}